\documentclass[twocolumn,trackchanges]{aastex701}
\usepackage{natbib}
\usepackage{comment}
\usepackage{amsmath}
\begin{document}

\title{Testing General Relativity with Binary Lens Microlensing Events}

\author{Xinyang Jin}
\affiliation{Institute for Frontiers in Astronomy and Astrophysics, Beijing Normal University, Beijing 102206, China}
\affiliation{School of Physics and Astronomy, Beijing Normal University, Beijing 100875, China}
\email{202521101104@mail.bnu.edu.com}  

\author{Shuo Cao}
\altaffiliation{\href{mailto:caoshuo@bnu.edu.cn}{caoshuo@bnu.edu.cn}}
\affiliation{Institute for Frontiers in Astronomy and Astrophysics, Beijing Normal University, Beijing 102206, China}
\affiliation{School of Physics and Astronomy, Beijing Normal University, Beijing 100875, China}
\email{caoshuo@bnu.edu.cn }

\author{Xuanni Chen}
\affiliation{School of Physics and Astronomy, Beijing Normal University, Beijing 100875, China}
\email{202411030042@mail.bnu.edu.cn}

\author{Yalong Nan}
\affiliation{Institute for Frontiers in Astronomy and Astrophysics, Beijing Normal University, Beijing 102206, China}
\affiliation{School of Physics and Astronomy, Beijing Normal University, Beijing 100875, China}
\email{202531101079@mail.bnu.edu.cn}

\author{Xuchen Tai}
\affiliation{Institute for Frontiers in Astronomy and Astrophysics, Beijing Normal University, Beijing 102206, China}
\affiliation{School of Physics and Astronomy, Beijing Normal University, Beijing 100875, China}
\email{202521101115@mail.bnu.edu.cn}

\author{Shuxun Tian}
\altaffiliation{\href{mailto:tshuxun@bnu.edu.cn}{tshuxun@bnu.edu.cn}}
\affiliation{School of Physics and Astronomy, Beijing Normal University, Beijing 100875, China}
\email{tshuxun@bnu.edu.cn}

\begin{abstract}

Einstein’s General Relativity (GR), which has been precisely tested on galaxy and solar system scales, is still poorly constrained in extremely weak gravitational fields (with dimensionless Newtonian potentials of $\Phi\sim10^{-8}$). In this Letter, we propose a new method to probe such a weak-field regime of gravity and measure the spatial curvature generated per unit mass. Specifically, binary microlensing events provide a laboratory to measure the parameterized post-Newtonian (PPN) parameter ($\gamma$), by combining the lensing mass ($M_{\mathrm{lens}}^{{\mathrm{GR}}}$) inferred from microlensing with the dynamical mass ($M_{\mathrm{dyn}}$) determined from interferometric astrometry. Astrometric tracking of binary lenses was previously hindered by their faintness and minuscule angular separations (typically a few milliarcseconds). This limitation has been alleviated by the deployment of GRAVITY+, which pushes the instrument's sensitivity and astrometric precision to unprecedented heights. Based on astrometric simulations of the binary microlensing event ASASSN-22av, we directly estimate $\gamma$ at the precision of $\sim 23\%$. Such test of GR in the extremely weak-field regime can be further improved to much higher precision ($\sim 3\%$), based on more binary microlensing events with precise determination of the dynamical/lens mass.


\end{abstract}



\section{INTRODUCTION} 

\begin{figure}[htbp]
\centering
\includegraphics[width=1.0\columnwidth]{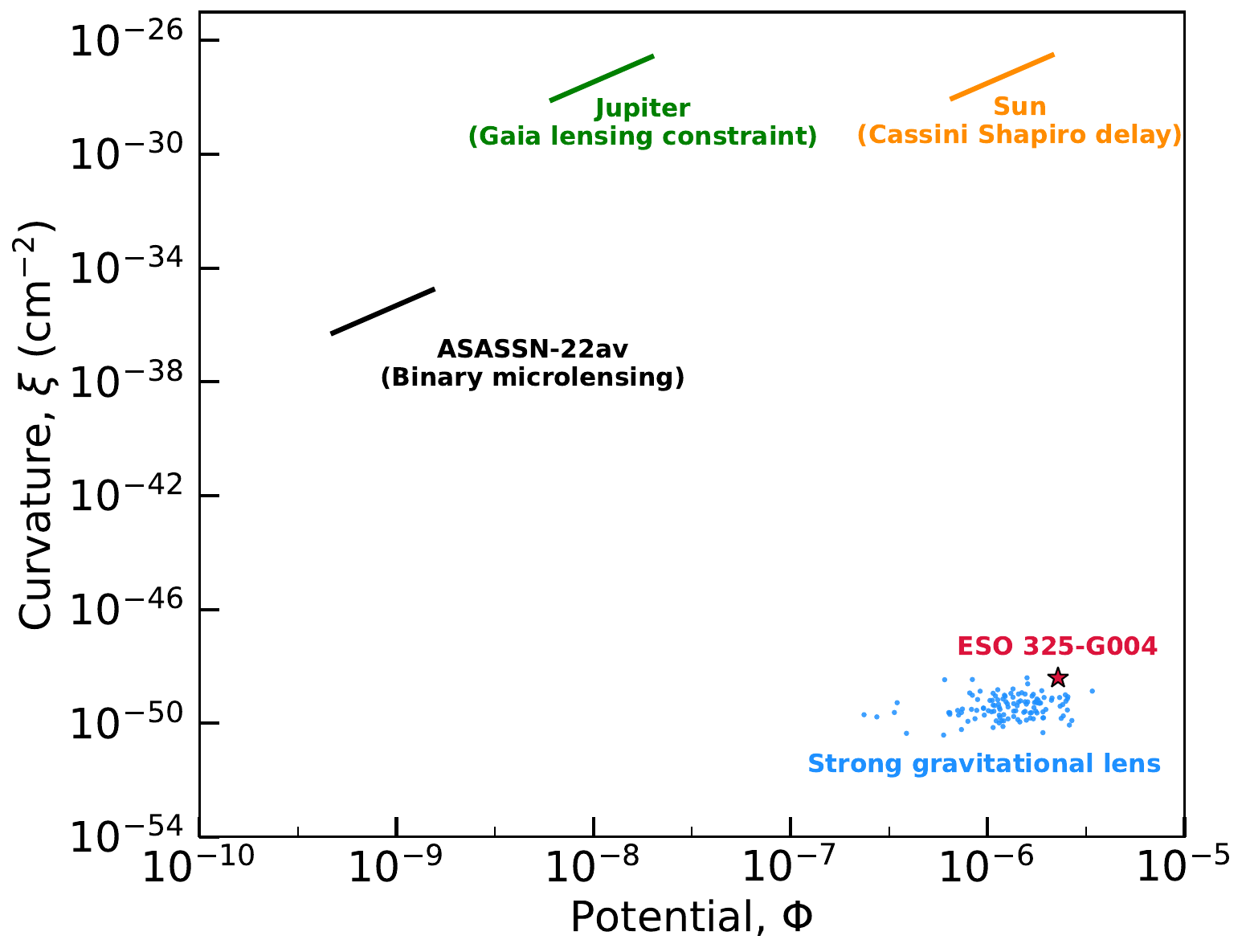}
\caption{Potential--curvature parameter space of various lensing-relevant systems, along with the Cassini mission. The horizontal axis shows the dimensionless Newtonian potential ($\Phi$), and the vertical axis represents the curvature ($\xi$) derived from the Kretschmann scalar\footnote{For the Schwarzschild metric, the Kretschmann scalar is expressed as $\xi = (R^{\alpha\beta\gamma\delta}R_{\alpha\beta\gamma\delta})^{1/2} = \sqrt{48} \frac{GM}{r^3 c^2}$, where $R$ represents the Riemann curvature tensor, following \cite{baker2015}.}. The line segments represent the values of $\Phi$ and $\xi$ calculated at grazing incidence along the ray-tracing path that contributes 95\% of the total deflection angle\footnote{Since the deflection angle comes from a long path, and different positions along this path have different gravitational environments, we only keep the part that makes the largest contribution to the deflection angle. Specifically, for Jupiter, we calculate the light rays that graze its surface, while using Gaia's high-precision astrometric data to detect light deflection usually requires considering the gravitational fields of the Sun or even Saturn \citep{jupiter}.}.  The black line denotes the microlensing event ASASSN-22av \citep{ASASSN-22av} adopted in our subsequent simulations, which also serves as a representative example of a typical Galactic M-dwarf lens. The blue dots show the distribution of strong-lensing galaxies from \citet{cao2015}, while the pentagram marks a precisely modeled galaxy, ESO 325-G004 \citep{Collett}.}
\label{fig:potential_curvature}
\end{figure}

General Relativity (GR), which stands as the cornerstone of modern cosmology, has withstood decades of rigorous experimental tests \citep{GRAVITY_TEST2,gravity_test} and remains remarkably robust. Although GR has been extraordinarily successful, it reveals its limitations on cosmological scales, most notably through the puzzles of dark matter and dark energy \citep{rotate,Accelerating_Universe}, as well as the modification of Einstein field equations \citep{clifton2012}.
To systematically compare these modified gravity theories, the Parameterized Post-Newtonian (PPN) framework was developed \citep{ppn}. In the weak-field limit, the PPN formalism parameterizes the departures from GR, allowing any metric theory of gravity to be described by a unified set of coefficients. In particular, the parameter $\gamma$, which quantifies the amount of spatial curvature produced per unit rest mass, is equal to unity ($\gamma=1$) in GR.
In previous studies, the constraints on the PPN parameter come primarily from two classical tests: light deflection and Shapiro time delay \citep{Shapiro}. 
Tests of $\gamma$ have covered vast spatial domains, spanning from strong gravitational lensing at galactic scales \citep{strong_lens,2026EPJC...86..650G} down to the precision Cassini tracking within the solar system \citep{cassini}. Nevertheless, it is essential to investigate potential deviations from GR in the extremely weak-field regime, characterized by a gravitational potential of $\Phi \sim 10^{-8}$. This regime represents a critical window for testing new physics beyond GR. For instance, Modified Newtonian Dynamics \citep[MOND;][]{MOND} alters the gravitational law precisely in the weak-field or low-acceleration limit as an alternative to dark matter. Furthermore, theories with screening mechanisms \citep[e.g.,][]{chameleon} suppress deviations in strong fields, indicating that recognizable departures from GR only emerge at large radii or within extremely diffuse environments. The positions of various probes for testing $\gamma$ are displayed within the potential-curvature diagram \citep{baker2015} in Fig.~\ref{fig:potential_curvature}.

In this work, we propose that binary microlensing \citep{double_lens} offers an alternative path to constrain $\gamma$ in the extremely weak-field regime (Fig.~\ref{fig:potential_curvature}), by combining the mass $M_{\mathrm{lens}}^{\mathrm{GR}}$ inferred through microlensing under GR with the Keplerian dynamical mass $M_{\mathrm{dyn}}$ obtained from astrometric observations once the source and lens are well resolved, where $M_{\mathrm{lens}}^{\mathrm{GR}}$ is interpreted in modified gravity theories as $\frac{1+\gamma}{2} M_{\mathrm{dyn}}$. This approach provides a testing ground independent of solar-system tests, enabling experimental verification across diverse interstellar environments due to the widespread galactic distribution of microlenses. Furthermore, unlike strong gravitational lensing, this method does not rely on complex galactic mass models. Conducting high-precision astrometry on binary microlensing events has long been difficult. However, this observational landscape is rapidly evolving with the upgrade of GRAVITY and the Very Large Telescope Interferometer \citep[VLTI;][]{vlti} in GRAVITY+ \citep{gravity+}. Leveraging its milliarcsecond-level angular resolution, researchers have already succeeded in resolving microlensing images \citep[e.g.,][]{dong,Przemek}. The Letter is organized as follows. Sec.~\ref{sec:binary_microlensing} starts from the lens equation to systematically derive the model parameters constrained by light curves, and evaluates how different approximations for the binary orbital motion impact the inferred $M_{\mathrm{lens}}^{\mathrm{GR}}$, determining whether $\gamma$ can be correctly recovered. In Sec.~\ref{Astrometric}, we simulate VLTI observations of ASASSN-22av \citep{ASASSN-22av}—the first binary microlensing event with multiple resolved images, which itself provides a highly precise measurement of $M_{\mathrm{lens}}^{\mathrm{GR}}$—to infer its $M_{\mathrm{dyn}}$. In Sec.~\ref{sec:results}, we discuss our results and future prospects.

\setcounter{footnote}{0}

\section{Binary microlensing} \label{sec:binary_microlensing}

In the weak-field limit, the spacetime metric can be described by two dimensionless potentials: the Newtonian potential $\Phi$ and the curvature potential $\Psi$. It is expressed as
\begin{equation}
ds^2 = -(1 + 2\Phi)dt^2 + (1 - 2\Psi)\delta_{ij}dx^i dx^j.
\label{eq:metric}
\end{equation}
$\Phi$ is associated with gravitational time dilation, while $\Psi$ characterizes the spatial curvature produced by gravity, with the PPN parameter defined through $\Psi = \gamma \Phi$. A value of $\gamma=1$ signifies the recovery of GR.

A microlensing event \citep{einstein,paczunski} occurs when a background star (i.e., the source), a foreground object (i.e., the lens), and the observer are closely aligned along the line of sight, observed as a time-dependent magnification of the source brightness throughout the event. During this period, the source is mapped into multiple images on the lens plane. Specifically, a light ray (or photon) which passes the lens at a distance $r$ is deflected by an angle of \citep{Will1993}
\begin{equation}
\hat{\alpha}(r) = \frac{2GM}{c^2 r}(1 + \gamma).
\label{eq:alpha}
\end{equation}
To describe the image positions, the angular separations on the lens plane are typically normalized by the Einstein angular radius $\theta_{\mathrm{E}}$, which in the PPN formalism is written as \citep{2010ApJ...708..750S}
\begin{equation}
\theta_E = \sqrt{\frac{4GM}{c^2} \frac{D_{\mathrm{LS}}}{D_{\mathrm{S}} D_{\mathrm{L}}} \frac{1+\gamma}{2}},
\label{eq:theta_E}
\end{equation}
where $D_{\mathrm{L}}$, $D_{\mathrm{S}}$, and $D_{\mathrm{LS}}$ denote the observer--lens, observer--source, and lens--source distances, respectively, and $M$ is the total mass of the lens.  
For a microlensing event where the impact parameter is on the order of the Einstein radius (i.e., $r = \theta_{\mathrm{E}} D_{\mathrm{L}}$), the scale of this localized potential is about
\begin{equation}
\Phi_{\mathrm{micro}} \sim \frac{GM}{c^2 \theta_{\mathrm{E}} D_{\mathrm{L}}} \approx 10^{-8},
\end{equation}
which places the lens system in the extremely weak-field regime. Since microlensing involves a complex, multi-parameter fitting process, we analyze below from a modeling perspective whether any implicit dependence on $\gamma$ is introduced into the expression of $\theta_{\mathrm{E}}$ during the fitting procedure (which would influence the expression of $M$).

The lens equation is given by \citep{lensequation}
\begin{equation}
\boldsymbol{y} = \boldsymbol{x} - \boldsymbol{\hat{\alpha}}(\boldsymbol{x}),
\label{eq:lens_base}
\end{equation}
where the dimensionless quantities are defined as $\boldsymbol{x} = \boldsymbol{\theta}/\theta_{\mathrm{E}}$ and $\boldsymbol{y} = \boldsymbol{\beta}/\theta_{\mathrm{E}}$, with $\boldsymbol{\theta}$ and $\boldsymbol{\beta}$ representing the angular positions of the image and the source, respectively. For a binary lens system, we can further express this equation in a complex formalism \citep{Witt1990}
\begin{equation}
y = x - \frac{m_1}{\bar{x}-\bar{x}_1} - \frac{m_2}{\bar{x}-\bar{x}_2}.
\label{eq:lens_complex}
\end{equation}
Here, $x_1$ and $x_2$ denote the positions of the binary lenses, and $\bar{x}$ is the complex conjugate of $x$. The normalized masses of the primary and secondary lenses are denoted by $m_1$ and $m_2$, with the mass ratio defined as $q = m_2/m_1$. Since the choice of the coordinate system does not affect the physical outcomes, for convenience, we position the lenses on the real axis such that $x_1 = s/2$ and $x_2 = -s/2$, where $s$ represents the projected separation between the lenses. We can then rewrite Eq. (\ref{eq:lens_complex}) as
\begin{equation}
y = x - \frac{1}{1+q}\frac{1}{\bar{x}-s/2} - \frac{q}{1+q}\frac{1}{\bar{x}+s/2}.
\label{eq:lens_rw}
\end{equation}

Note that Eq. \eqref{eq:lens_rw} can be expressed as a fifth-order polynomial in $x$ \citep{Witt1990,Witt1995}: $\sum_{i=0}^5 c_i x^i = 0$. The magnification $\mu$ is written as 
\begin{equation}
\mu = \sum_{i=1}^5 1/\left.\left|1 - \frac{\partial y}{\partial \bar{x}}\overline{\frac{\partial y}{\partial x}}\right|\right|_{x=x_i}. 
\end{equation}
As $\mu$ is an observable, combining the equations above yields $f(y,q,s)=0$, which serves as an abstract notation representing the algebraic relation after eliminating the image positions $x$. Considering that the source is not always a perfect point and that $y$ varies continuously over time, the light-curve model can be parameterized as
\begin{equation}
f(y(u_0, t_0, t_{\mathrm{E}}), q, s) = 0,
\label{eq:light_curve}
\end{equation}
where $u_0$ is the dimensionless impact parameter (the closest approach of the source trajectory to the coordinate origin), $t_0$ is the time of peak magnification usually expressed as HJD, and $t_{\mathrm{E}}$ is the Einstein radius crossing time \citep{paczunski}.
For long-timescale microlensing events, the orbital motion of the Earth induces an observable parallax, which is defined as
\begin{equation}
\label{eq:pi_E}
\boldsymbol{\pi}_{\mathrm{E}} \equiv \frac{\pi_{\mathrm{rel}}}{\theta_{\mathrm{E}}}\frac{\boldsymbol{\mu}_{\mathrm{rel}}}{\mu_{\mathrm{rel}}},
\end{equation}
with $\pi_{\mathrm{rel}} \equiv \frac{\mathrm{1 AU}}{D_{\mathrm{L}}} - \frac{\mathrm{1 AU}}{D_{\mathrm{S}}}$ and $\boldsymbol{\mu}_{\mathrm{rel}}$ denoting the lens-source relative proper motion vector \citep{gould2000,gould2004}. These parameters describe the trajectory of the source center as a complex position $y(t)$ in the lens plane
\begin{equation}
y(t) = \left( \frac{t-t_0}{t_{\mathrm{E}}} + \boldsymbol{\pi}_{\mathrm{E}} \cdot \boldsymbol{\Delta s} \right) + i \left( u_0 + \boldsymbol{\pi}_{\mathrm{E}} \times \boldsymbol{\Delta s} \right),
\label{eq:trajectory}
\end{equation}
where $\boldsymbol{\Delta s}$ represents the projected positional offset of the Earth relative to the Sun. In practice, $\boldsymbol{\pi}_{\mathrm{E}}$ is decomposed into the north and east components in equatorial coordinates: $(\pi_{{\mathrm{E}},{\mathrm{N}}}, \pi_{{\mathrm{E}},{\mathrm{E}}})$. 

In general, the source cannot be treated as a point mass, as it possesses a finite angular size. The finite-source parameter is defined as $\rho \equiv \theta_*/\theta_{\mathrm{E}}$ \citep{rho}, where $\theta_*$ is the angular radius of the source projected onto the lens plane. To account for the orientation of the binary axis in equatorial coordinates, we introduce the position angle $\alpha$, which defines the angle of the source trajectory relative to the binary axis. Consequently, Eq.~(\ref{eq:light_curve}) can be reformulated as
\begin{equation}
f(y(u_0, t_0, t_{\mathrm{E}}, \theta_{\mathrm{E}}, \pi_{{\mathrm{E}},{\mathrm{N}}}, \pi_{{\mathrm{E}},{\mathrm{E}}}, \alpha), q, s)= 0
\label{eq:light_curve_parallax},
\end{equation}
where we have replaced $\rho$ with $\theta_{\mathrm{E}}$.

Now the mass inferred from a microlensing event under the assumption of GR is \citep{2012JCAP...03..016C}
\begin{equation}
M_{\mathrm{lens}}^{\mathrm{GR}} = \frac{c^2 \mathrm{AU}}{4G}\frac{\theta_{\mathrm{E}}}{\pi_{\mathrm{E}}}.
\label{eq:mass_microlensing_GR}
\end{equation}
Introducing $M_{\mathrm{dyn}}$ as the mass determined through independent dynamical measurements -- which rely on Keplerian motion and depend only on $\Phi$ -- one obtains \citep{strong_lens}
\begin{equation}
M_{\mathrm{lens}}^{\mathrm{GR}}=\frac{1+\gamma}{2} M_{\mathrm{dyn}}.
\label{eq:mass_lens_dyn}
\end{equation}
Note that $M_{\mathrm{lens}}^{\mathrm{GR}}$ here serves purely as a numerical quantity—namely, the nominal mass inferred under the standard GR assumption—which, in the context of modified gravity, is interpreted as the expression on the right-hand side. Since $M_{\mathrm{dyn}}$ probes only $\Phi$ while $M_{\mathrm{lens}}^{\mathrm{GR}}$ responds to $(\Phi+\Psi)$, combining these two mass measurements directly determines $\gamma$.

Eq. (\ref{eq:light_curve_parallax}) represents the relation satisfied by a static binary lens. However, when the binary rotates through a small angle during the event, the light curve is generally characterized by a linear motion approximation to the static model \citep{line_motion} $s = s_0 + \dot{s}(t-t_0)$ and $\alpha = \alpha_0 + \dot{\alpha}(t-t_0)$. We now examine whether Keplerian equations enter the previous parametric framework, as their inclusion would allow $M_{\mathrm{dyn}}$ to enter the microlensing model, potentially mixing $M_{\mathrm{dyn}}$ with $M_{\mathrm{lens}}^{{\mathrm{GR}}}$. For both static and linear-motion binary lenses, the lens Eq. (\ref{eq:lens_rw}) is dimensionless and therefore independent of the value of $\gamma$. Moreover, since $\gamma$ does not appear in the subsequent equations, the modeling procedure is entirely independent of kinematic equations.

If linear motion in the lens plane is insufficient to characterize the light curves (e.g., when the event timescale occupies a significant fraction of the orbital period), the line-of-sight relative motion should be incorporated, described by $s_z = s_{z0} + \dot{s}_z(t - t_0)$ \citep{full_orbit}. So far, this is the complete description of the system, from which all standard Keplerian orbital elements can be calculated.
In full-orbit modeling, the microlensing parameters specify the instantaneous three-dimensional relative position ($\mathbf{r}$) and velocity ($\mathbf{v}$) of the binary components at a reference epoch. These phase-space coordinates uniquely determine the specific orbital energy ($E$) and specific angular momentum ($\mathbf{h}$), from which the corresponding orbital elements can be derived. The resulting orbit is then propagated to each observation epoch to obtain the projected binary relative position, enabling a more accurate calculation of the microlensing magnification for fitting the observed light curve. However, the mass governing $E$ and $\mathbf{h}$ is expressed as $M_{\mathrm{lens}}^{\mathrm{GR}}$,\footnote{In the traditional microlensing framework, the total mass is treated as $M_{\mathrm{lens}}^{\mathrm{GR}}$. The orbital dynamics are governed by the specific orbital energy $E = -GM/2a$ and the specific angular momentum vector $\mathbf{h} = \mathbf{r} \times \mathbf{v}$, where $G$ is the gravitational constant, $M$ represents the total lens mass, $a$ is the semi-major axis, $\mathbf{r}$ denotes the relative position vector, and $\mathbf{v}$ signifies the relative velocity vector.} rendering these events inapplicable under the current framework. To use these events, we need to modify the existing fitting codes to include a new free parameter $\gamma$ and constrain it using Eq.~\eqref{eq:mass_lens_dyn}.

In this work, we do not consider microlensing events that require full-orbit modeling, although full-orbit modeling is theoretically viable. 
If the existing modeling pipelines are modified in the future as described above, the sample of microlensing events capable of testing $\gamma$ could be significantly expanded. In fact, microlensing events that require full-orbit modeling are typically closer to us and exhibit higher brightness, which is important for high-precision tests of $\gamma$. In order to derive constraints on $\gamma$, we first select cases where $M_{\mathrm{lens}}^{\mathrm{GR}}$ has been determined with high precision, and then filter these cases that are suitable for measuring $M_{\mathrm{dyn}}$ via astrometric methods. Although these events are currently rare, this situation is expected to change with the launch of the Roman Space Telescope \citep[RST;][]{roman}.

\begin{table*}[t]
\centering
\caption{Best-fit parameters for the binary lens event ASASSN-22av.}
\label{tab:lens_parameters}

\setlength{\tabcolsep}{8pt}
\renewcommand{\arraystretch}{1.4}

\hspace{-2cm}
\begin{tabular}{cccccccc}

\hline
\hline

$M_1$ ($M_{\odot}$) 
& $M_2$ ($M_{\odot}$) 
& $r_{\perp}$ ($\mathrm{au}$) 
& $D_{\mathrm{L}}$ ($\mathrm{kpc}$) 
& $\theta_{\mathrm{E}}$ ($\mathrm{mas}$) 
& $\pi_{\mathrm{E}}$ 
& $q$ 
& $\Psi$ ($\mathrm{deg}$) \\

\hline

$0.256_{-0.006}^{+0.008}$ 
& $0.130_{-0.007}^{+0.007}$ 
& $6.83_{-0.27}^{+0.31}$ 
& $2.29_{-0.08}^{+0.08}$ 
& $0.724 \pm 0.012$ 
& $0.35_{-0.01}^{+0.02}$ 
& $0.51 \pm 0.03$ 
& $259.4_{-0.2}^{+0.2}$ \\

\hline
\hline

\end{tabular}

\end{table*}

\begin{table*}[ht!]
\centering
\caption{Injected parameters and the best-fit orbital elements retrieved from two distinct mock observational scenarios: ASPRO2 simulations and a simple Gaussian noise control. \textbf{Values for $M_{\mathrm{dyn}}$ and $\pi_{\mathrm{L}}$ are derived from the multi-parameter ($P, a, \pi_{\mathrm{E}}, \pi_{\mathrm{S}}, q, \theta_{\mathrm{E}}$) joint sampling.}}
\label{tab:fit_results}

\begin{tabular}{l@{\hspace{0.6cm}}ccccccc}
\hline
\hline

Configuration & $M_{\mathrm{dyn}}$ ($M_\odot$) & $P$ (yr) & $\pi_{\mathrm{L}}$ & $a$ (mas) & $i$ ($^\circ$) & $\omega$ ($^\circ$) & $\Omega$ ($^\circ$) \\

\hline

Injected Values 
& $0.386$ 
& $28$ 
& $0.437$ 
& $2.93$ 
& $60.0$ 
& $46.0$ 
& $50.74$ \\

ASPRO2 Mock Data
& $0.377^{+0.053}_{-0.047}$ 
& $27.976^{+0.072}_{-0.072}$ 
& $0.441^{+0.020}_{-0.019}$ 
& $2.94^{+0.01}_{-0.01}$ 
& $60.1^{+0.7}_{-0.7}$ 
& $226.9^{+0.9}_{-0.9}$ 
& $230.5^{+0.6}_{-0.6}$ \\

Gaussian Control
& $0.383^{+0.080}_{-0.067}$ 
& $28.120^{+0.766}_{-0.766}$ 
& $0.4408^{+0.020}_{-0.019}$ 
& $2.96^{+0.12}_{-0.12}$ 
& $59.0^{+3.1}_{-3.1}$ 
& $48.6^{+6.8}_{-6.8}$ 
& $50.4^{+3.6}_{-3.6}$ \\

\hline
\end{tabular}

\tablecomments{Due to the lack of radial velocity (RV) data, the posterior distribution exhibits the well-known $(\omega, \Omega)$ mirror degeneracy. The MCMC chains for the ASPRO2 mock data and the Gaussian control happen to converge to the two degenerate solutions, which are physically equivalent in pure astrometric modeling.}

\end{table*}

\section{VLTI Astrometric Simulation}
\label{Astrometric}

To determine $M_{\mathrm{dyn}}$, astrometric methods are required to track the relative positions of binary stars at different epochs. Note that although the combination of radial velocity (RV) measurements with microlensing datasets could improve the constraints on some parameters, the binary masses can not be resolved by RV data alone \citep{2016ApJ...821..121Y}. Such limitation arises from the spectroscopic mass function of
{
\begin{equation}
\mathcal{M} = \frac{(m_2 \sin i)^3}{(m_1 + m_2)^2},\end{equation}}considering the degeneracy between the individual masses ($m_1$, $m_2$) and the orbital inclination ($i$). On the other hand, the orbital elements are determined astrometrically, enabling the total mass of the binary to be derived from Kepler's Third Law. The mass ratio $q$ is independent of the parameter $\gamma$, allowing us to resolve the individual masses of the components.

Since the lens systems are located at large distances, the angular separation of the binary components is always only a few milliarcseconds (mas), requiring angular resolutions beyond the capabilities of conventional single-aperture telescopes. Instead, long-baseline optical interferometry can provide angular resolutions at this level. Currently, the most powerful near-infrared interferometric facility is the VLTI at the European Southern Observatory (ESO), which can combine either four 8-meter Unit Telescopes (UTs) or four movable 1.8-meter Auxiliary Telescopes (ATs). The VLTI has successfully tracked and analyzed the orbits of faint stars around $\text{Sgr~A}^*$ \citep{gravity}, demonstrating its capability for high-precision astrometry and making it a unique facility for microlensing binary measurements.

\subsection{Simulated Observing Strategy}

As a case in point, we consider the binary microlensing event ASASSN-22av, whose $M_{\mathrm{lens}}^{\mathrm{GR}}$ value has been determined with high precision via the joint modeling of photometric and VLTI interferometric data. This system consists of two M-dwarfs with an angular separation of $\sim 3$ mas and a total $K$-band magnitude of $\sim 18.5$. A subset of the best-fit parameters adopted from the microlensing solution of \cite{ASASSN-22av} is summarized in Table~\ref{tab:lens_parameters}. To evaluate the precision of the $M_{\mathrm{dyn}}$ measurement, we performed a simulation of this binary event based on the technical capabilities of GRAVITY+.

In the first step, we defined a set of fixed elliptical orbital parameters. Owing to the lack of complete orbital information in the microlensing solution, we adopt an illustrative assumption, setting the eccentricity $e$ and inclination $i$ to typical values for binary stars. Furthermore, we assumed that the binary system was at apastron during the microlensing event. Combined with the binary position angle derived from the lensing parameters, this allowed us to calculate the argument of periastron $\omega$ and the longitude of ascending node $\Omega$, respectively. We compared the fitted orbital parameters against these reference values, with the detailed results compiled in Table~\ref{tab:fit_results}. Subsequently, we employed the ASPRO2 \citep{aspro2} to estimate the uncertainties in the squared visibility amplitudes ($|V|^2$) and closure phases ($T_3\phi$). We employed the single-beam astrometry method \citep{2022A&A...657L..12G}, as both components of the binary lens resided within the same interferometric field of view (IFOV). The lens separation was extracted by fitting a binary model to the $|V|^2$ and $T_3\phi$. Considering that precise modeling of the stellar structure is not required, we adopted the low spectral-resolution, combined-polarization mode in this analysis. The simulations assumed "Good sky conditions" (seeing = $0.7''$, $t_0$ = $4.4\text{ ms}$, $h_0$ = $5250.0\text{ m}$). The simulated scheduling began in the year 2070, with an observational cadence of three years. For each epoch, the hour angle spanned from -1 h to 1 h, and the exposure consisted of 11 individual frames with a $30\text{ s}$ integration time. This configuration yielded an average signal-to-noise ratio (SNR) of approximately 2.8 for $|V|^2$. The source was designated as the Fringe Tracking \citep[FT;][]{ft1,ft2} star because its $K$-band is $9.27\text{ mag}$,  well above the FT brightness requirement. Both binary components were modelled as point sources, which is a physically sound assumption given their distance of over $2\text{kpc}$. We developed an ancillary script to calculate the relative angular positions from the elliptical orbital elements at each observational epoch, which were then fed into the model.

Two distinct approaches were considered for data processing. The first approach extracts the $u\text{-}v$ coordinates along with the noiseless $|V|^2$ and $T_3\phi$ observables from the simulated data, subsequently injecting anticipated measurement uncertainties into these idealized datasets. However, this approach assigns one or several fixed error values across different $u\text{-}v$ coordinates and wavelengths, which compromises the reliability of the final estimates. The second approach leverages the data generated by ASPRO2, which incorporate both statistical noise and instrumental biases, and back-calculates the relative positions from these synthetic observables. We adopted this latter scheme. Since the noise model in ASPRO2 intrinsically accounts for statistical uncertainties, we generated 10 independent datasets for each epoch, with each dataset fitted independently using the same binary model and optimization procedure.

Given that the microlensing event occurred in January 2022, we assumed that the barycenter of the binary lens coincided with the source at this specific epoch. The relative proper motion of merely $3\text{ mas/yr}$ between the lens and the source makes their angular separation remain within the IFOV over recent years. Despite the Gaussian attenuation of the source brightness imposed by the single-mode fiber \citep{2017}, the residual flux of the source could still contaminate the interference fringes of the binary lens. We centered the primary star (defined as the more massive component) within the IFOV (or the photocenter if the binary is unresolved). Based on the model by \citet{Wang2021}, by the year 2070 -- when the separation reaches approximately $144\text{ mas}$ -- the single-mode fiber coupling efficiency drops to approximately $1.6 \times 10^{-8}$. At this stage, the source's contribution to the interference fringes becomes negligible, while its baseline flux remains sufficient to serve as an effective FT guide star.

Before generating the simulated data, it is crucial to determine the flux ratio of the binary components, as it strongly influences $|V|^2$ and $T_3\phi$. However, due to the extremely tight angular separation of microlensing binaries, almost all photometric surveys fit the combined flux of the binary lens as a single unresolved object, making it difficult to extract flux ratios. Guided by the empirical mass-luminosity relation for M-dwarfs, we adopted a flux ratio of 5:1 between the primary and secondary stars. We note that this assumption induces a potential risk: if the secondary component is very faint, the $|V|^2$ and $T_3\phi$ signals will become indistinguishable from those of a single point-like lens. Under such scenarios, an alternative strategy should be deployed. By performing the phase-referenced dual-field relative astrometry between the primary (or photocenter) and the source, and subsequently subtracting the proper motion effects, we can detect the subtle motion of the primary component around the barycenter.

\begin{figure}[ht!]  
\centering          
\includegraphics[width=1.0\columnwidth] {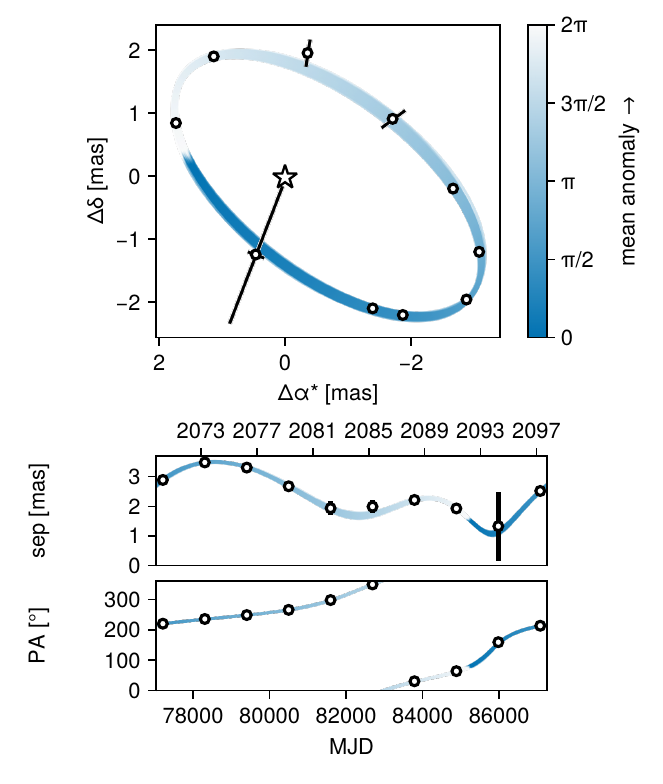} 
\caption{The simulated observational baseline starts from the year 2070, at which epoch the background source flux contamination within the instrument's IFOV becomes negligible. Data points without visible error bars have uncertainties that are smaller than the symbol size on this scale. The blue curve represents the possible orbits, with the color bar on the right indicating the mean anomaly at different phases. To mimic a realistic observing campaign given the expected long orbital period of the binary lens system, data points are sampled at a regular cadence of three years, extending until 2097.}
\label{fig:orbit_fit}
\end{figure}

\subsection{Orbit Reconstruction}
We utilized PMOIRED\footnote{\url{https://github.com/amerand/PMOIRED}}\citep{PMOIRED} to fit the relative angular positions of the binary system based on the simulated interferometric observables. To mitigate sensitivity to local minima, we adopted a multi-stage optimization framework comprising a coarse grid search over the companion's position space, a local grid search, and a final gradient-based refinement to simultaneously fit both astrometric and photometric parameters. To enhance convergence stability and reduce parameter degeneracies, the flux ratio was held fixed during the coarse and fine grid search phases. Once the companion's position was closely constrained, the flux ratio was allowed to vary and fitted jointly with the astrometric parameters in the final optimization step. The final outputs for each epoch were determined as the mean values of the fitted relative positions across the 10 noise realizations. The associated uncertainties were derived from the standard deviation of the parameter distribution, while the corresponding covariance matrices were evaluated from the sample covariance.

Using the recovered relative positions, we employed Octofitter\footnote{\url{https://github.com/sefffal/Octofitter.jl}} to infer the orbital parameters of the binary. Although Octofitter was originally developed for exoplanetary systems, its relative orbit formalism remains valid for stellar binaries. As detailed in the appendix of the foundational paper \citep{Octofitter2024}, the kinematic equations still hold for stellar binaries, with the sole exception of Equation~(A18), which adopts a negligible-mass approximation for the secondary component.


\section{Results and discussion}
\label{sec:results}

We present the simulated relative orbital trajectory of the secondary component around the primary on the plane of the sky in Fig.~\ref{fig:orbit_fit}, where the 2094 data point shows a high uncertainty. This directly reflects the physical limitations of the observation, arising from the exceptionally close angular separation of the binary system at this epoch. Given that the bias at this specific epoch remains relatively minor, we retain this data point in our Markov Chain Monte Carlo (MCMC) fitting pipeline, as the likelihood function naturally assigns it a lower weight, preventing it from dominating the final orbital solution. The root-mean-square errors (RMSEs) of the residuals along the $\Delta\alpha^*$ (Right Ascension, incorporating the $\cos\delta$ factor) and $\Delta\delta$ (Declination) directions are $22.6\,\mu\mathrm{as}$ and $98\,\mu\mathrm{as}$, respectively, while the mean formal uncertainties are $\bar{\sigma}_{\alpha^*} = 89\,\mu\mathrm{as}$ and $\bar{\sigma}_\delta = 178\,\mu\mathrm{as}$. Without the 2094 epoch, the RMSEs become $23.8\,\mu\mathrm{as}$ and $83.5\,\mu\mathrm{as}$, with $\bar{\sigma}_{\alpha^*} = 49.6\,\mu\mathrm{as}$ and $\bar{\sigma}_\delta = 76.3\,\mu\mathrm{as}$. The fact that $\bar{\sigma}_\delta$ is significantly larger than $\bar{\sigma}_{\alpha^*}$ stems directly from the geometry of the UTs array at the scheduled observation dates (selected around April 10th annually), where the maximum effective baseline along the $v$-axis is relatively short, resulting in sparser spatial frequency sampling compared to the $u$-axis \citep{2017}. For comparison, we also performed a simulation by adding a Gaussian noise component obeying $N(0, 200\,\mu\mathrm{as})$ to the theoretical positions in both $\Delta\alpha^*$ and $\Delta\delta$. Although this approach is highly simplified, it allows us to effectively assess the precision of the orbital parameters under a larger error regime. Table~\ref{tab:fit_results} details the initial parameters alongside the fitted values obtained from the interferometric framework and the Gaussian noise simulation.

\begin{figure}[ht!]  
\centering          
\includegraphics[width=1.0\columnwidth]{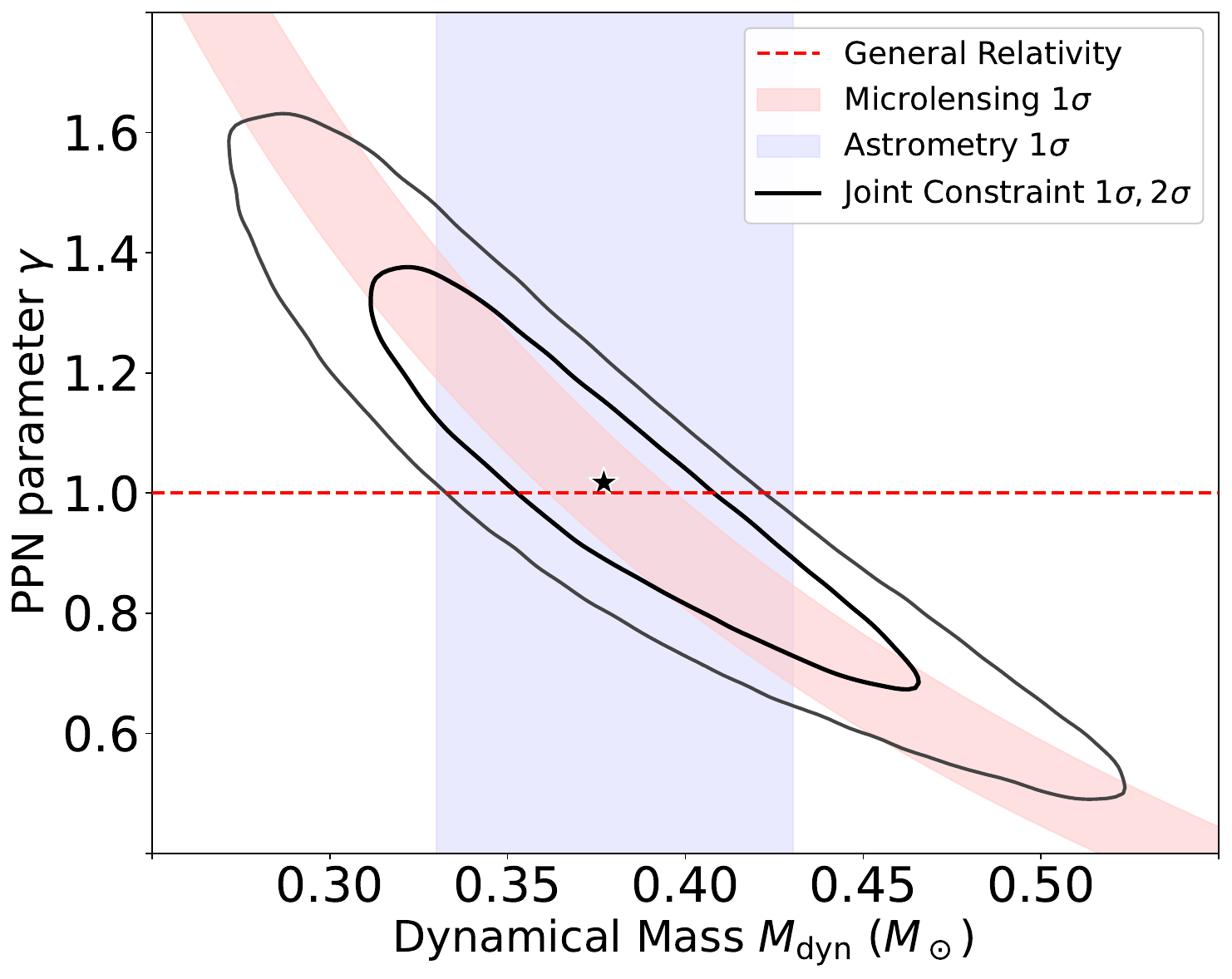}
\caption{Joint posterior distribution of $M_{\mathrm{dyn}}$ and $\gamma$. The purple shaded region denotes the $1\sigma$ confidence interval for the simulated $M_{\mathrm{dyn}}$, while the pink shaded region represents the corresponding $1\sigma$ interval for $M_{\mathrm{lens}}^{\mathrm{GR}}$. The solid black contours denote the $1\sigma$ and $2\sigma$ joint credible regions.}
\label{fig:joint_posterior}
\end{figure}

 One should note that $M_{\mathrm{dyn}}$ is dependent on $\pi_{\mathrm{L}}^{-3}$ (where $\pi_{\mathrm{L}} = \pi_{\mathrm{S}} + \theta_{\mathrm{E}}\pi_{\mathrm{E}}$ denotes the lens parallax), while $M_{\text{lens}}^{\text{GR}}$ scales with $\theta_{\text{E}} / \pi_{\text{E}}$. Therefore, it is necessary to account for the covariance between the two mass estimates when the adopted $\pi_{\text{L}}$ is derived from a lens solution rather than measured independently. Specifically, in the absence of the original MCMC chains from the lens fit, we draw samples for $\pi_{\text{E}}$, $\theta_{\text{E}}$, and $q$ from independent normal distributions. Note that the $\theta_{\text{E}}$ reported in \cite{ASASSN-22av} corresponds to the primary lens; thus, we incorporate the factor $(1+q)$ to estimate $M_{\text{lens}}^{\text{GR}}$. By combining these with the independently measured $\pi_{\text{S}}$ and the orbital elements ($a$ (mas) and $P$) extracted from the mock astrometric MCMC chains, we construct the joint posterior distribution of $\gamma$ and $M_{\text{dyn}}$ in Fig.~\ref{fig:joint_posterior}. A joint analysis combining astrometry and microlensing yields $\gamma = 1.017^{+0.244}_{-0.224}$, corresponding to a relative precision of $\sim23\%$.

Precise measurement of $M_{\mathrm{dyn}}$ plays an important role in determining $\gamma$. Assuming $P$ is well-constrained by the long-term observational baseline, the error propagation yields
\begin{equation}
    \frac{\sigma_{M_{\mathrm{dyn}}}}{M_{\mathrm{dyn}}} \approx 3 \sqrt{ \left(\frac{\sigma_a}{a}\right)^2 + \left(\frac{\sigma_{\pi_{\mathrm{L}}}}{\pi_{\mathrm{L}}}\right)^2 }.
    \label{eq:error_prop}
\end{equation}
As presented by the ASPRO2 mock analysis in Table~\ref{tab:fit_results}, the relative precision of $a$ is tightly constrained to $\sigma_a/a \approx 0.4\%$. In contrast, the amplified $\pi_{\mathrm{L}}$ term $\sigma_{\pi_{\mathrm{L}}}/\pi_{\mathrm{L}} \approx 4.4\%$ heavily dominates the total error budget. Consequently, more precise $\pi_{\mathrm{L}}$ measurements in the future are expected to yield a corresponding improvement in the $\gamma$ constraint. Apart from deploying astrometric methods to determine $M_{\mathrm{dyn}}$, it may also become feasible in the future to constrain $\gamma$ in a few cases by directly measuring the asteroseismic masses \citep[e.g.,][]{asteroseismology} of lenses. However, the lenses are rarely giant stars with pronounced asteroseismic signals. Instead, they are predominantly solar-like stars or M-dwarfs \citep{mass}, whose typical radial oscillation amplitudes are on the order of merely $\mathrm{cm\,s}^{-1}$ \citep{Amplitudes}. This imposes exceedingly stringent requirements on both next-generation instrumental sensitivity and pristine target selection criteria.

To obtain an accurate $M_{\mathrm{lens}}^{\mathrm{GR}}$, acquiring high-precision microlensing light curves is paramount, as this directly dictates the extraction precision of $\pi_{\mathrm{E}}$ and $\rho$ (see Eq.~(\ref{eq:mass_microlensing_GR})). Simultaneously, deriving a precise $M_{\mathrm{dyn}}$ requires not only high-quality astrometric data but also a highly accurate source parallax ($\pi_{\mathrm{S}}$), which is geometrically coupled with $\pi_{\mathrm{E}}$ and $\pi_{\mathrm{L}}$. This presents a stringent requirement for current microlensing samples: it demands that the lens should not be too faint, and that the binary components must have adequate angular separation. In the era of Roman Space Telescope \citep[RST;][]{roman}, it is anticipated to detect 27,000 microlensing events with $|u_0| < 1$ over the course of its mission \citep{roman}. A fraction of $5\%$ events is expected to be 
binary lens events with distinct caustic crossings \citep{2019BAAS...51c.211G}. We further assume that $5\%$ of such subsample satisfies the observational requirements in this analysis, 
leaving a robust catalog of $\sim 68$ golden events. Considering the precision is inversely proportional to $\sqrt{N}$ (where $N$ is the number of systems), such a test of GR in the extremely weak-field regime can be further improved to a precision of $\sim 3\%$.
Nevertheless, by synergizing space-based observatories (e.g., the RST, Spitzer \citep{zang}, and Gaia \citep{Gaia}) with ground-based networks (such as Optical Gravitational Lensing Experiment \citep[OGLE;][]{ogle1,ogle2}, and VLTI), an increasing number of high-precision samples suitable for this rigorous test will be obtained in the near future. If we could directly utilize events with full-orbit solutions derived from high-precision light curves, the situation would become much more straightforward. As detailed in Sec.~\ref{sec:binary_microlensing}, by treating $\gamma$ as a free parameter within the fitting routine, it might be possible to constrain $\gamma$ without relying on any astrometric data. Prominent examples of such ideal targets include Gaia16aye \citep{Gaia16aye} and Gaia20fnr \citep{Gaia20fnr}.

\begin{acknowledgments}

This work is supported by Beijing Natural Science Foundation (No. 1242021), National Natural Science Foundation of China (Nos. 12433001; 12405050), and the Fundamental Research Funds for Central Universities.

\end{acknowledgments}

\bibliography{microlens}
\bibliographystyle{aasjournalv7}

\end{document}